\documentclass[12pt]{article}
\usepackage[dvipdfmx]{graphicx}
\catcode`\@=11
\@addtoreset{equation}{section}

\global\arraycolsep=2pt
\usepackage{amsbsy,amssymb,latexsym,amsfonts,amsmath}
\usepackage{graphicx,color}

\allowdisplaybreaks

\begin{document}

\begin{flushright}
\parbox{4.2cm}
{RUP-19-31}
\end{flushright}

\vspace*{0.7cm}

\begin{center}
{ \large Exclusion Inside or at the Border of Conformal Bootstrap Continent}
\vspace*{1.5cm}\\
{Yu Nakayama}
\end{center}
\vspace*{1.0cm}
\begin{center}

Department of Physics, Rikkyo University, Toshima, Tokyo 171-8501, Japan

\vspace{3.8cm}
\end{center}

\begin{abstract}
How large can anomalous dimensions be in conformal field theories? What can we do to attain larger values? One attempt to obtain large anomalous dimensions efficiently is to use the Pauli exclusion principle. Certain operators constructed out of constituent fermions cannot form bound states without introducing non-trivial excitations. To assess the efficiency of this mechanism, we compare them with the numerical conformal bootstrap bound as well as with other interacting field theory examples. In two-dimensions, it turns out to be  the most efficient: it saturates the bound and is located at the (second) kink. In higher dimensions, it more or less saturates the bound but it may be slightly inside.

\end{abstract}

\thispagestyle{empty} 

\setcounter{page}{0}

\newpage

\section{Introduction}
It is a remarkable fact that in two-dimensions, the critical Ising model is equivalent to a free massless Majorana fermion. This results in an unexpected selection rule of the operator product expansion (OPE) for the energy operator $\epsilon$
\begin{align}
\epsilon \times \epsilon = 1 + 0 \cdot \epsilon + \epsilon' + \cdots \ , 
\end{align}
where the coefficient in front of $\epsilon$ does not vanish, say, in the three-dimensional critical Ising model, but it does vanish in two-dimensions.  

In the Landau-Ginzburg viewpoint, the OPE can be interpreted as
\begin{align}
\phi^2 \times \phi^2 = 1 + 0 \cdot \phi^2 + \phi^4 + \cdots \ .
\end{align}
This selection rule does not arise from the Lagrangian symmetry of the $\phi^4$ theory and it is very hard to predict e.g. by using $\epsilon$ expansions. It is more than vanishing of this one term. The Virasoro symmetry further tells us that infinitely many Virasoro descendant operators of $\phi^2$ have zero OPE coefficients unlike the ones in higher dimensions. 

To see how this happens, one may resort to the Virasoro symmetry and study the Virasoro conformal bootstrap, but there is a more intuitive understanding of this phenomenon. It is simply related to the free fermion representation. 
If we accept that the energy operator $\epsilon$ is given by a bilinear of free massless Majorana fermion $\epsilon \sim \psi_L \psi_R$, we can immediately understand its origin as a chiral $\tilde{\mathbb{Z}}_2$ symmetry acting on the Majorana fermion (i.e. $\psi_L \to -\psi_L$, $\psi_R \to \psi_R$). 

Another feature here is that while we are working with the {\it free} fermion, we can still see a strongly coupled nature of the critical $\phi^4$ theory. The conformal dimension of the first operator $\epsilon'$ that appears in the OPE of $\epsilon \time \epsilon = 1 + \epsilon' + \cdots$ is not twice  that of  $\Delta_\epsilon = 1 $. It is rather $\Delta_{\epsilon'} = 4$. The free fermion viewpoint gives an interesting interpretation of this strongly coupled nature: it is the Pauli exclusion principle that makes $\psi_L \psi_L$ vanish so that the first available term is $\epsilon' = T\bar{T} = \psi_L \partial \psi_L \psi_R \bar{\partial}\psi_R$ (rather than $\psi_L \psi_L \psi_R \psi_R$ = 0). 

It is an interesting question to ask how large values we can attain in anomalous dimensions of general conformal field theories (CFTs). The large anomalous dimensions can be utilized to find concrete examples to solve hierarchy problems in various fine-tunings (notably in Higgs sector of the standard model of particle physics) or to find examples of self-organized criticality.\footnote{For conformal bootstrap approaches to these problems, see e.g. \cite{Poland:2018epd} and reference therein.} In the context of holography, it is related to how large interactions we can introduce without violating quantum gravity constraint. 

As we have alluded above, use of the Pauli exclusion principle seems an efficient way to obtain large anomalous dimensions. In this paper, we would like to assess how efficient this mechanism can be. As a benchmark, we would like to compare it with the numerical conformal bootstrap bound. In two-dimensions, we will show that the Pauli exclusion is the most efficient. The four-point functions out of free Majorana fermion bilinear will saturate the bound and it is located at the (second) kink of the numerical conformal bootstrap bound. In higher dimensions, it more or less saturates the bound but it may be slightly inside the allowed conformal bootstrap continent.



\section{Method}
In order to assess the efficiency of the Pauli exclusion principle to obtain large anomalous dimensions in conformal field theories, we compare it with the bound coming from the numerical conformal bootstrap \cite{Poland:2018epd}. For our purpose, we study the crossing symmetry of four-point functions among identical scalar operators with assumed global symmetries. With the unitarity, the crossing symmetry constraint can be reduced to a (infinite dimensional) semi-definite problem that can be investigated numerically \cite{Rattazzi:2008pe}\cite{Poland:2010wg}\cite{Rattazzi:2010yc}\cite{Vichi:2011ux}\cite{Poland:2011ey}\cite{ElShowk:2012ht}\cite{ElShowk:2012hu}\cite{Kos:2013tga}\cite{El-Showk:2013nia}\cite{El-Showk:2014dwa} \cite{Nakayama:2014lva}\cite{Nakayama:2014yia}\cite{Bae:2014hia}\cite{Chester:2014gqa}\cite{Iha:2016ppj}\cite{Nakayama:2016knq}\cite{Nakayama:2017vdd}\cite{Stergiou:2019dcv}\cite{Go:2019lke}. 

Our goal in this paper is to study general bounds rather than making a precise prediction for a particular target conformal field theory e.g. the three-dimensional critical Ising model. While bootstrapping mixed correlation functions has been a very powerful tool for the latter purpose \cite{Kos:2014bka}\cite{Kos:2015mba}\cite{Iliesiu:2015qra}\cite{Iliesiu:2015qra}\cite{Nakayama:2016jhq}\cite{Kos:2016ysd}\cite{Li:2016wdp}\cite{Go:2019lke}\cite{Kousvos:2019hgc}\cite{Chester:2019ifh}, we will only use a bound from a single correlation function. This is partly because we empirically know that a simple application of mixed correlation functions do not give a better bound for larger dimensions that we will be interested in.  

We will study the four-point functions of identical scalar operators in fundamental representations of $O(N)$ global symmetry with $N=1,2,3$. Here $O(1) = \mathbb{Z}_2$. For the $\mathbb{Z}_2$ case, the relevant OPE sum rule is
\begin{align}
0 = \sum_{O \in \Phi \times \Phi} \lambda^2_O F \ 
\end{align}
while for the $O(N>1)$ case, the sum rules  are given by 
\begin{align}
0 = \sum_{S^+} \lambda^2_{S^+} \left(
    \begin{array}{c}
      0 \\
      F  \\
      H 
    \end{array}
  \right) + \sum_{T^+} \lambda^2_{T^+}  \left(
    \begin{array}{c}
      F \\
      \left(1-\frac{2}{N}\right) F  \\
      -\left(1+\frac{2}{N} \right)H
    \end{array}
  \right) + \sum_{A^-} \lambda^2_{A^-}  \left(
    \begin{array}{c}
      -F \\
       F  \\
      -H 
    \end{array}
  \right) 
\end{align}
where $(\pm)$ denotes the even $(+)$ or odd $(-)$ spin contributions. We have used the convention
\begin{align}
F & = v^{\Delta_{\Phi}} g_{\Delta_O,l}(u,v) - u^{\Delta_{\Phi}} g_{\Delta_O,l}(v,u) \cr
H & = v^{\Delta_{\Phi}} g_{\Delta_O,l}(u,v) + u^{\Delta_{\Phi}} g_{\Delta_O,l}(v,u)
\end{align}
with the conformal block $g_{\Delta_O,l}$ being normalized as in \cite{Hogervorst:2013kva} (which can be explicitly found in \cite{Dolan:2003hv}). The unitarity assumes $\lambda^2_O \ge 0$ and $\Delta_O \ge d-2+l$.

Once we rewrite the constraint as a semi-definite program, one may use a numerical algorithm to solve the problem. We use cboot \cite{cboot} to generate the semi-definite program to be  solved by SDPB \cite{Simmons-Duffin:2015qma}\cite{Landry:2019qug}, but our results can be reproduced by other available software in the literature \cite{Paulos:2014vya}\cite{Behan:2016dtz}.

In the actual numerical conformal bootstrap analysis, we have to specify several cut-off parameters. The most important parameter for our purpose is the search space dimension of functionals to exclude given conformal data. It is specified $n+m \le \Lambda$, where $n$ and $m$ are number of $z$,$\bar{z}$ derivatives acting on the conformal blocks at the crossing symmetric point $z=\bar{z}=1/2$  (therefore the search space dimension scales as $\Lambda^2$).
Compared with the previous studies e.g. \cite{Kos:2013tga}, we take larger cut-offs (at the cost of search speed) because we are more interested in the bounds for the higher values of conformal dimensions where the conformal blocks are exponentially small and we need to increase the cut-offs to obtain non-trivial bounds. Accordingly, with the higher search space dimension, we have to increase the digits kept during the numerical computation. A typical parameter set can be found in Table \ref{table:1}.

\begin{table}[htbp]
\begin{tabular}{|c||c|c|}
\hline
$\Lambda$  & 41 for $O(2)$ & 55 for $\mathbb{Z}_2$  \\ \hline
 $\nu_{\mathrm{max}}$ & 45 & 59     \\
 spins & up to 52 & up to 60  \\
 precision  & 1024 & 1408 \\
 dualityGapThreshold & $10^{-70}$ & $10^{-80}$ \\
primalErrorThreshold & $10^{-100}$ &  $10^{-100}$ \\
dualErrorThreshold & $10^{-100}$ & $10^{-100}$ \\
initialMatrixScalePrimal & $10^{70}$ & $10^{70}$\\
initialMatrixScaleDual & $10^{70}$ & $10^{70}$  \\
maxComplementarity& $10^{200}$ & $10^{200}$ \\ 
\hline
\end{tabular}
	\caption{Sample parameter set used in our numerical conformal bootstrap. The parameter $\nu_{\mathrm{max}}$ is the number of poles (with respect to the conformal dimension) kept in approximating the conformal block.}
	\label{table:1}
\end{table}

\section{Results}

\subsection{Two dimensions}
In two-dimensions, Majorana-Weyl condition can be imposed on a spinor. Let us consider one free Majorana fermion $\psi_L$, $\psi_R$ with left and right chirality to construct a scalar operator $\epsilon = \psi_L\psi_R$ with conformal dimension $\Delta_\epsilon = 1$. This theory admits a chiral symmetry of $\psi_L \to -\psi_L$, $\psi_R \to \psi_R$ under which $\epsilon$ is odd. Since the Majorana-Weyl fermion shows the Pauli exclusion principle i.e. $\psi_L \psi_L =0$ because $\psi_L$ is one-component, the first operator appearing in the OPE of $\epsilon \times \epsilon$ is $T\bar{T} = \psi_L \bar{\partial} \psi_L \psi_R \partial \psi_R$ which has dimension $\Delta_{T\bar{T}} =4$, and it is much larger than $2\Delta_{\epsilon} = 2$.

\begin{figure}[htbp]
	\begin{center}
		\includegraphics[width=12.0cm,clip]{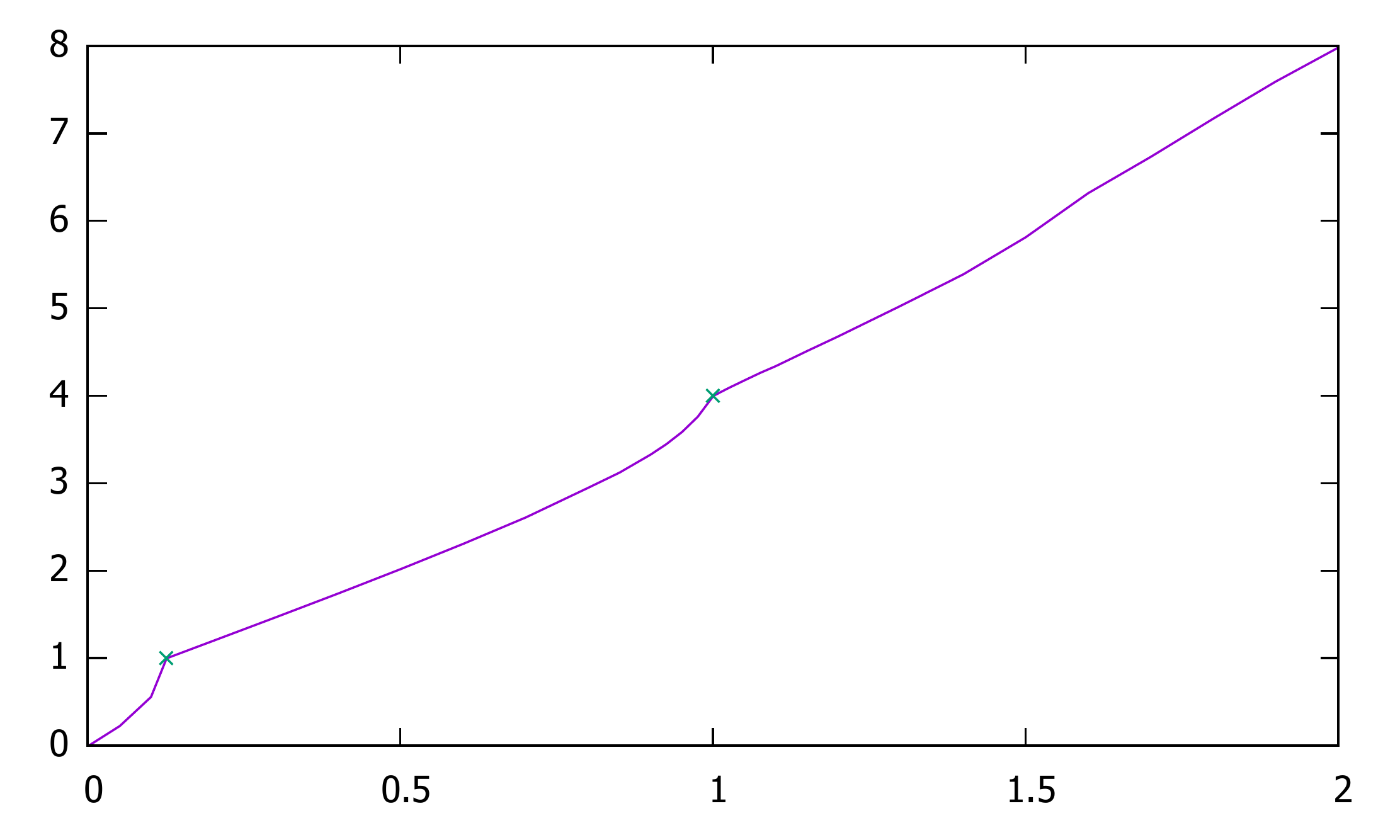}
	\end{center}
	\caption{Unitarity bound for $\Delta_{\text{even}}$ in two-dimensional $\mathbb{Z}_2$ symmetric CFT as a function of $\Delta_{\text{odd}}$. The point with green cross is the free massless Majorana fermion with the Pauli exclusion principle.}
	\label{fig:2d}
\end{figure}

Let us assess the efficiency of this mechanism by studying the numerical conformal bootstrap bound.
In Fig \ref{fig:2d}, we have presented the results of the numerical conformal bootstrap bound with $\mathbb{Z}_2$ symmetry. The bound shows two interesting features at $\Delta_{\mathrm{odd}} = 0.125$ and $\Delta_{\mathrm{odd}}=1.0$. The first one at $\Delta_{\mathrm{odd}}=0.125$ with $\Delta_{\mathrm{even}}=1.0$ was originally pointed out in \cite{ElShowk:2012ht} and it was identified with the two-dimensional critical Ising model or a free massless Majorana fermion with a spin operator $\sigma$ whose exact conformal dimension is $\Delta_{\sigma}=1/8$ and gives $\sigma \times \sigma = 1+\epsilon+\cdots$ with $\Delta_{\epsilon}=1$. The detailed analysis of spectrum out of the extremal functional saturating the bound at the first kink was done in \cite{El-Showk:2014dwa}, agreeing with the exact spectrum of a free massless Majorana fermion.

Here we would like to focus on the second kink at $\Delta_{\mathrm{odd}} = 1.0$ with $\Delta_{\mathrm{even}}=4.0$.\footnote{The appearance of the second kink can be found in recent studies \cite{Gowdigere:2018lxz}\cite{Paulos:2019fkw} as well. S.~Rychkov has informed the author that it was first observed by S.~El-Showk.} This is precisely the location of the free fermion OPE of $\epsilon \times \epsilon = 1 + T\bar{T} + \cdots$ mentioned above. This means that the Pauli exclusion principle in the two-dimensional conformal field theory gives the most efficient way to obtain the large anomalous dimension (at least at $\Delta_{\mathrm{odd}} = 1$). We also observe that the spectrum from the extremal functional is precisely what we expect in a free Majorana fermion.  For example, the estimated spectrum of the spin zero operator in the OPE is given by 4.00003, 8.0003, 12.006, 16.09 and 20.7 with no contribution from the descendant of $\epsilon$. 

We have a couple of comments about the numerical bound in Fig \ref{fig:2d}. Between $\Delta_{\mathrm{odd}} = 0.125$ and $\Delta_{\mathrm{odd}}=1.0$, we observe a non-linear functional form of the bound. In the literature, it was conjectured that the bound will become a straight line (in the $\Lambda \to \infty$ limit) of $\Delta_{\mathrm{even}} = \frac{8}{3}\Delta_{\mathrm{odd}}+\frac{2}{3}$ up to $\Delta = 0.5$, where the generalized minimal models are saturating the bound \cite{Liendo:2012hy}\cite{Paulos:2019gtx}. It is certain that the bound must be above the straight line, we have a reservation if the bound is precisely saturated by the generalized minimal models (given that the curve must be non-trivial at least between $\Delta_{\mathrm{odd}} = 0.5$ and $\Delta_{\mathrm{odd}} =1.0$). This possible non-saturation was as far as the author knows first claimed by T.~Ohtsuki.\footnote{In particular, the numerical conformal bootstrap analysis tells that the convergence of the bound at $\Delta_{\mathrm{odd}} =0.125$ and $\Delta_{\mathrm{odd}} = 1.0$ is much faster than at $\Delta_{\mathrm{odd}} = 0.5$ as we increase $\Lambda$.}

Beyond $\Delta = 1$, it seems that the numerical bound we obtained is slightly stronger than the free theory expectation of $\Delta = 4\Delta_{\epsilon}$ (as is claimed in \cite{Liendo:2012hy}). Note that in a free scalar theory or in massless Thirring model, there always exists a moduli operator that has dimension $2$, which is well inside the bound.

\subsection{Three dimensions}
In three dimensions, Majorana condition can be imposed on a spinor. Let us consier a (two-component) free massless Majorana fermion $\psi_\alpha$ with conformal dimension $\Delta_{\psi}=1$. The classical action of a free massless Majorana fermion is invariant under time-reversal $T$ but the Majorana mass term $\epsilon = \psi^\alpha \psi_{\alpha}$ with conformal dimension $\Delta_{\epsilon}=2$ is odd under $T$. From the viewpoint of the numerical conformal bootstrap among scalar operators, one may effectively regard $T$ as a global $\mathbb{Z}_2$ symmetry \cite{Poland:2018epd}. Thus, effectively we have the OPE of $\epsilon = \psi^\alpha \psi_{\alpha}$ as $\epsilon \times \epsilon = 1 + \epsilon' + \cdots$ without the appearance of $\epsilon$. 

Furthermore, the Pauli exclusion principle tells that $\epsilon'$ has an ``anomalous dimension" of $\Delta_{\epsilon'}=6$ rather than $2\Delta_{\epsilon} = 4$ because $(\psi^{\alpha} \psi_{\alpha})^2 = 0$.\footnote{In the first print of the author's book on the higher dimensional conformal field theory \cite{Nakayamabook}, there is a small misprint on this point.} Therefore one may expect, as in two dimensions, there can be a non-trivial feature in the conformal bootstrap bound with the $\mathbb{Z}_2$ symmetry.

\begin{figure}[htbp]
	\begin{center}
		\includegraphics[width=12.0cm,clip]{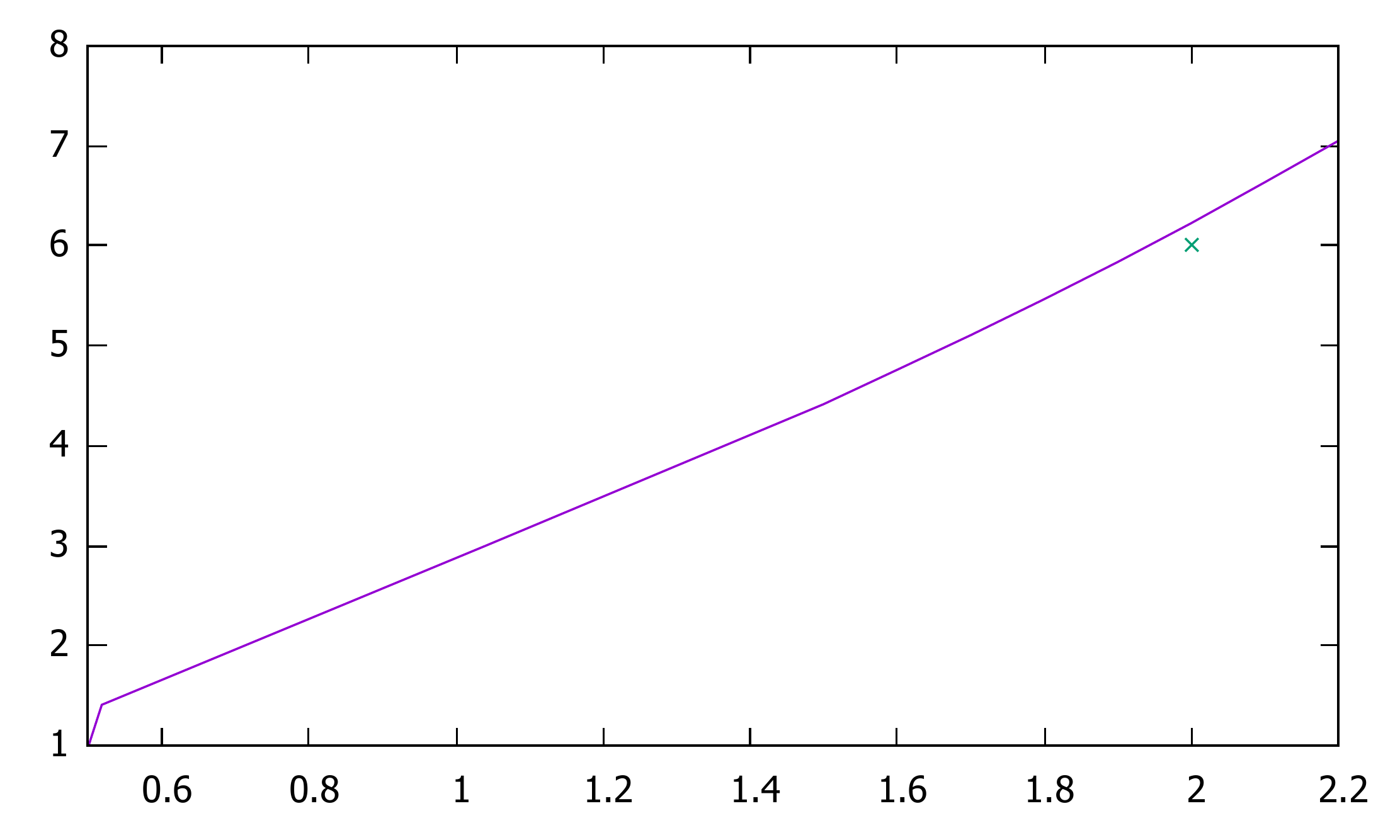}
	\end{center}
	\caption{Unitarity bound for $\Delta_{\text{even}}$ in three-dimensional $\mathbb{Z}_2$ symmetric CFT as a function of $\Delta_{\text{odd}}$. The point with green cross is the free Mahorana fermion with the Pauli exclusion principle.}
	\label{fig:3dz2}
\end{figure}

\begin{figure}[htbp]
	\begin{center}
		\includegraphics[width=12.0cm,clip]{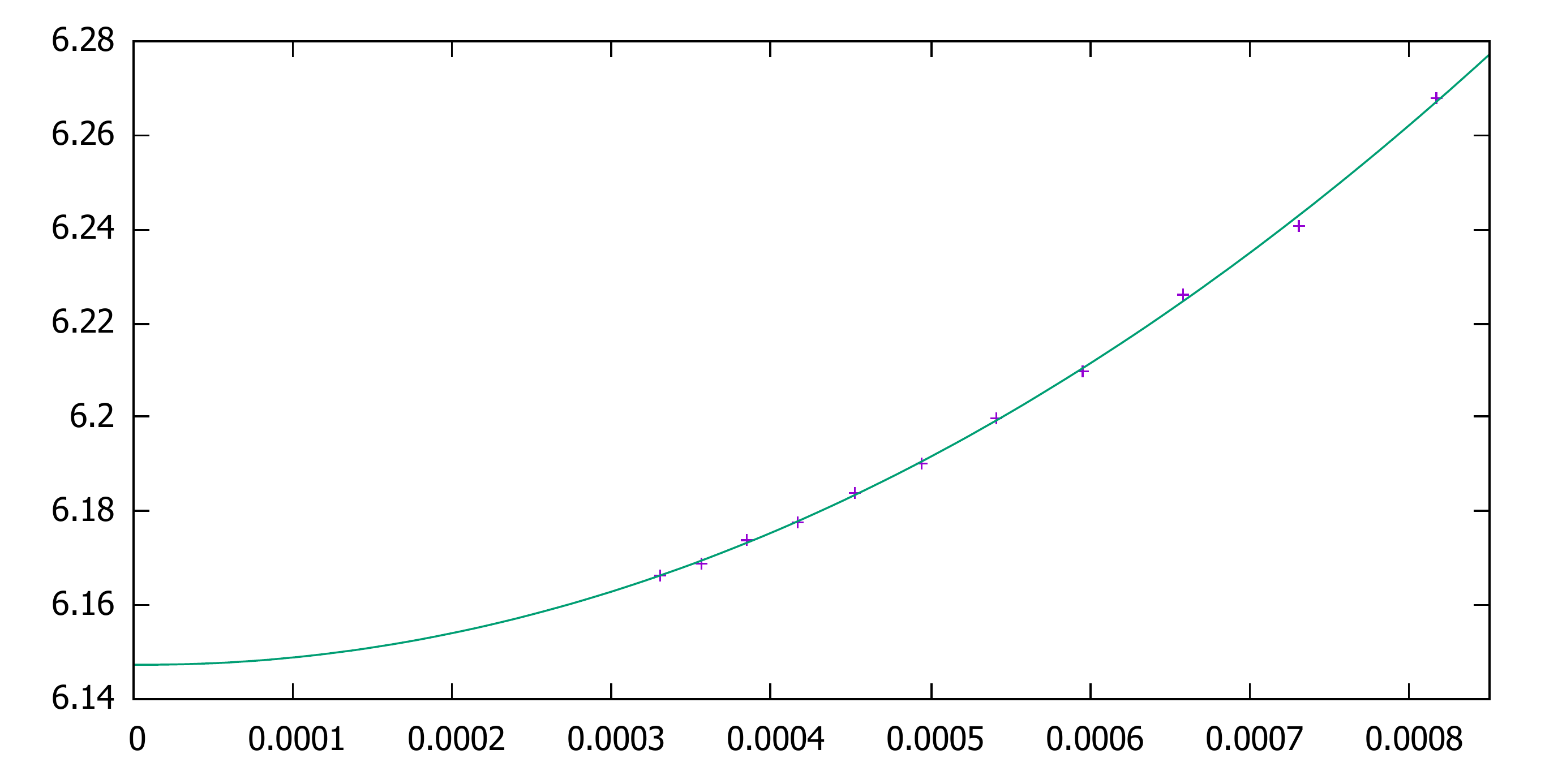}
	\end{center}
	\caption{The bound at $\Delta_{\mathrm{odd}}=2.0$ as a function of the cut-off parameter $1/\Lambda^2$ (between $\Lambda=35$ and $\Lambda=55$). The green curve is a numerical extrapolation.}
	\label{fig:3dconv}
\end{figure}

Fig. \ref{fig:3dz2} shows the numerical conformal bootstrap bound with $\Lambda =41$. It does not show the second kink unlike in two dimensions, but it might be the case that the OPE of a free Majorana fermion bilinear saturates the bound. In other words, we want to assess if the Pauli exclusion principle gives the most efficient way to obtain large anomalous dimensions. To see whether this is the case, we have to take the large $\Lambda$ limit. Although it is not conclusive, up to $\Lambda=55$ the bound at $\Delta_{\mathrm{odd}}=2.0$ does not seem to converge to $\Delta_{\mathrm{even}}=6.0$, while the low $\Lambda$ result looks promising (see Fig \ref{fig:3dconv}). Our crude extrapolation suggests $\Delta_{\mathrm{even}} \ge 6.147$.
We have also studied the spectrum but it does not show any characteristic integer behavior unlike in two dimensions. 

One observation here is that the OPE data of the free Majorana fermion $\epsilon = \psi^\alpha \psi_{\alpha}$ considered here satisfies all the conditions (i.e. there are no other relevant operators) applied in \cite{Kos:2014bka} to isolate the three-dimensional critical Ising model in numerical conformal bootstrap with mixed correlation functions. There it was observed that in addition to an island corresponding to the critical Ising model, there exists a vast continent that satisfies the condition. The free Majorana fermion does sit inside the continent albeit it may not sit at the border \cite{Nakayamabook}. 

Another interesting class of the Pauli exclusion principle is to consider free massless Dirac fermions with $U(1)$ or $SU(2)$ global symmetry. Let us take a massless Dirac fermion with $U(1)$ charge $1/2$, and we can form a charge $1$ scalar operator $\Phi = \psi_+^\alpha \psi_{+\alpha}$ with dimension $\Delta_{1} = 2$. Now the idea is that the charge $2$ operator constructed out of $\Phi \times \Phi$ shows the Pauli exclusion principle because $(\psi_+^\alpha\psi_{+\alpha})^2 = 0$. The lowest dimensional scalar operator in $\Phi \times \Phi$ has conformal dimension $\Delta_2 = 6$.

\begin{figure}[htbp]
	\begin{center}
		\includegraphics[width=12.0cm,clip]{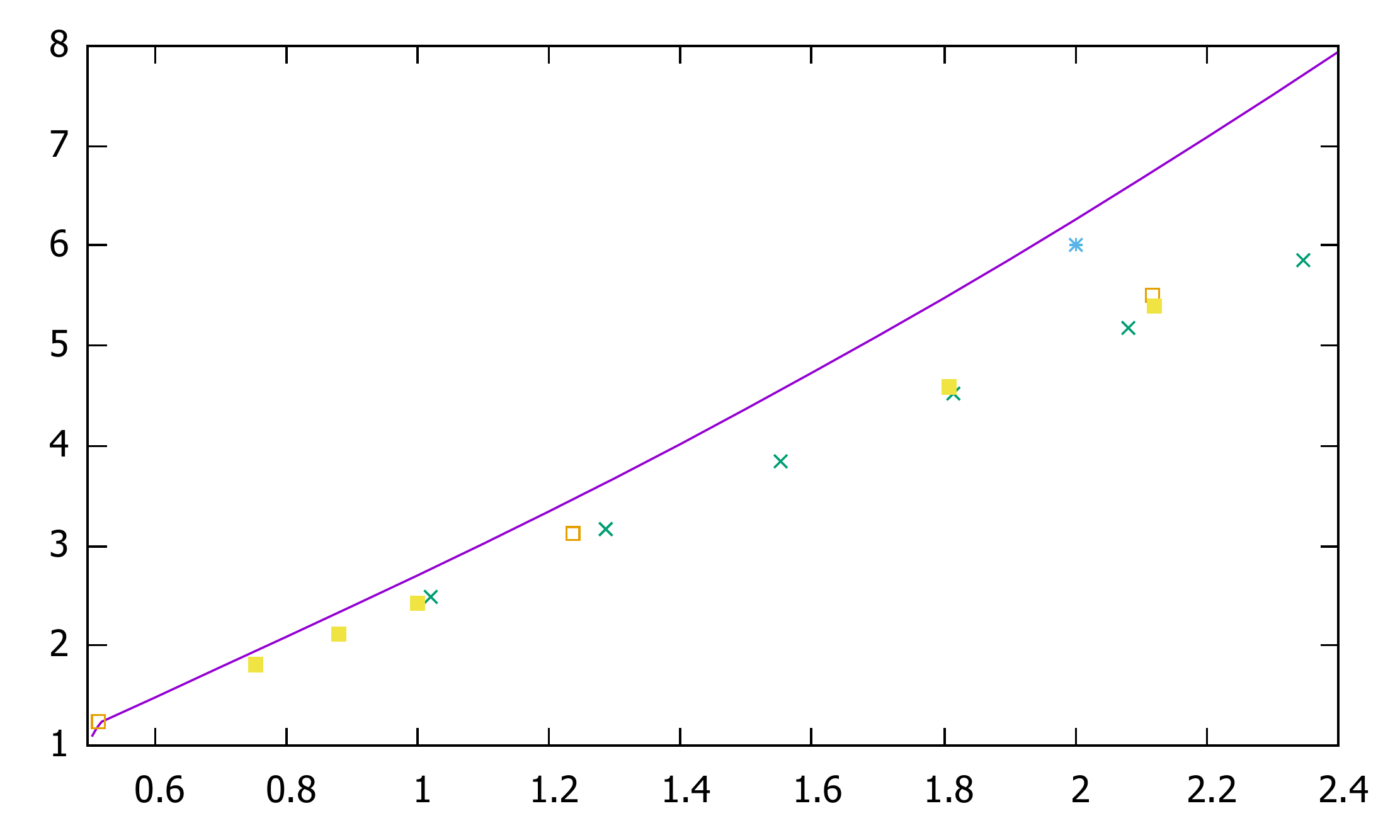}
	\end{center}
	\caption{Unitarity bound for $\Delta_{2}$ in three-dimensional $O(2)$ symmetric CFT as a function of $\Delta_{1}$. As a comparison, we have plotted the free massless Dirac fermion (blue asterisk
), critical $O(2)$ Landau-Ginzburg model (red white box), monopole operators in scalar QED (yellow box), and monopole operators in QED (green cross).}
	\label{fig:3do2}
\end{figure}

In order to assess the efficiency of the mechanism, we show the numerical conformal  bootstrap bound of $\Delta_2$ as a function of $\Delta_1$ in Fig \ref{fig:3do2}. Again, it is not immediately obvious if the bound is saturated by the free massless Dirac fermion in the infinite $\Lambda$ limit, but we do see that it is close to the bound. For comparison, we also plotted various other interacting CFTs with $U(1)$ symmetry computed by various methods \cite{Banerjee:2017fcx}\cite{Pufu:2013vpa}\cite{Dyer:2015zha}. We see that the Pauli exclusion principle is much more efficient than these interacting examples.

One related question is if the bound becomes a straight line with the slope $2\sqrt{2}$ in the infinite $\Lambda$ limit. This limit corresponds to the asymptotic  growth of conformal dimensions of large charge $Q$ operators in generic $U(1)$ symmetric CFTs in three dimensions: $\Delta_Q \sim Q^{3/2}$ \cite{Hellerman:2015nra}\cite{Jafferis:2017zna}. For the free fermion case this comes from the very simple scaling argument: the total energy scales as $E = \int^{p_F} d^{d-1}p |p| \propto p_F^d$ while the total charge scale as $Q = \int^{p_F} d^{d-1} p \propto p_F^{d-1}$ with respect to the Fermi energy $p_F$. If the series of operators with charge $2^n$ saturates the bound, the slope will be asymptotically $2\sqrt{2}$. In this sense it is nothing but the consequence of the Pauli exclusion principle.

The current situation inferred from our numerical bound is that the slope is slightly larger than $2\sqrt{2}$, but with more precision and with larger $\Delta_1$, it may become closer to this value or it may be systematically larger for the other reasons (but it cannot be smaller than $2\sqrt{2}$ anyway). We realize that the interacting examples in Fig \ref{fig:3do2} are more or less on the line with slope $2\sqrt{2}$ even though most of the charges are small. 
Note, however, that the slope below $\Delta_1 =2.0$ cannot be as small as $2\sqrt{2}$ because then one would exclude the free Dirac fermion $(\Delta_1,\Delta_2) = (2,6)$, so it is indeed the problem of larger $\Delta_1$ which becomes more and more difficult to obtain in the current numerical method because we need more precision and larger cut-offs. 

Similarly, let us take two massless Majorana fermions and form a $SU(2)$ doublet. Then one may construct the $SU(2)$ triplet $O^a = \psi_i^\alpha \sigma_{ij}^a \psi_{j \alpha}$ with spin zero. Again the Pauli exclusion principle demands that spin zero operator with the symmetric traceless representation in $O^a \times O^a$ OPE has non-trivial ``anomalous dimension" of $\Delta_T=6$ for $\Delta_{f} = 2$. Let us briefly compare it with the numerical conformal bootstrap bound. At $\Lambda = 41$ the bound is $\Delta_T \ge 6.253$ for $\Delta_f=2.0$. As may be expected in the previous studies for smaller $\Delta_f$, the bound is slightly better than the $U(1)$ case (for a fixed $\Lambda$), but the free fermion may not yet saturate the bound.

\subsection{Four dimensions}
In four-dimensions, Majorana condition or Weyl condition can be imposed on a spinor. Unlike in three dimensions, we do not have a global symmetry acting on a single Majorana mass term, so we do not have an analogue of $\mathbb{Z}_2$ symmetry with the Pauli exclusion principle. If we have a massless Weyl fermion or two massless Majorana fermions, one may use chiral $U(1)$ or $SU(2)$ symmetry to generate ``anomalous" dimensions from the Pauli exclusion principle. The situation then is very close to the one studied in the previous subsection. The Weyl fermion bilinear $(\psi^\alpha\psi_{\alpha})$ has conformal dimension $\Delta_1 = 3$ with a unit charge under  chiral $U(1)$, but $(\psi^\alpha\psi_{\alpha})^2 =0$ so that the lowest dimension operator in the symmetric representation (i.e. charge two operator) must have $\Delta_2 = 8$ rather than $2\Delta_1 = 6$.

\begin{figure}[htbp]
	\begin{center}
		\includegraphics[width=12.0cm,clip]{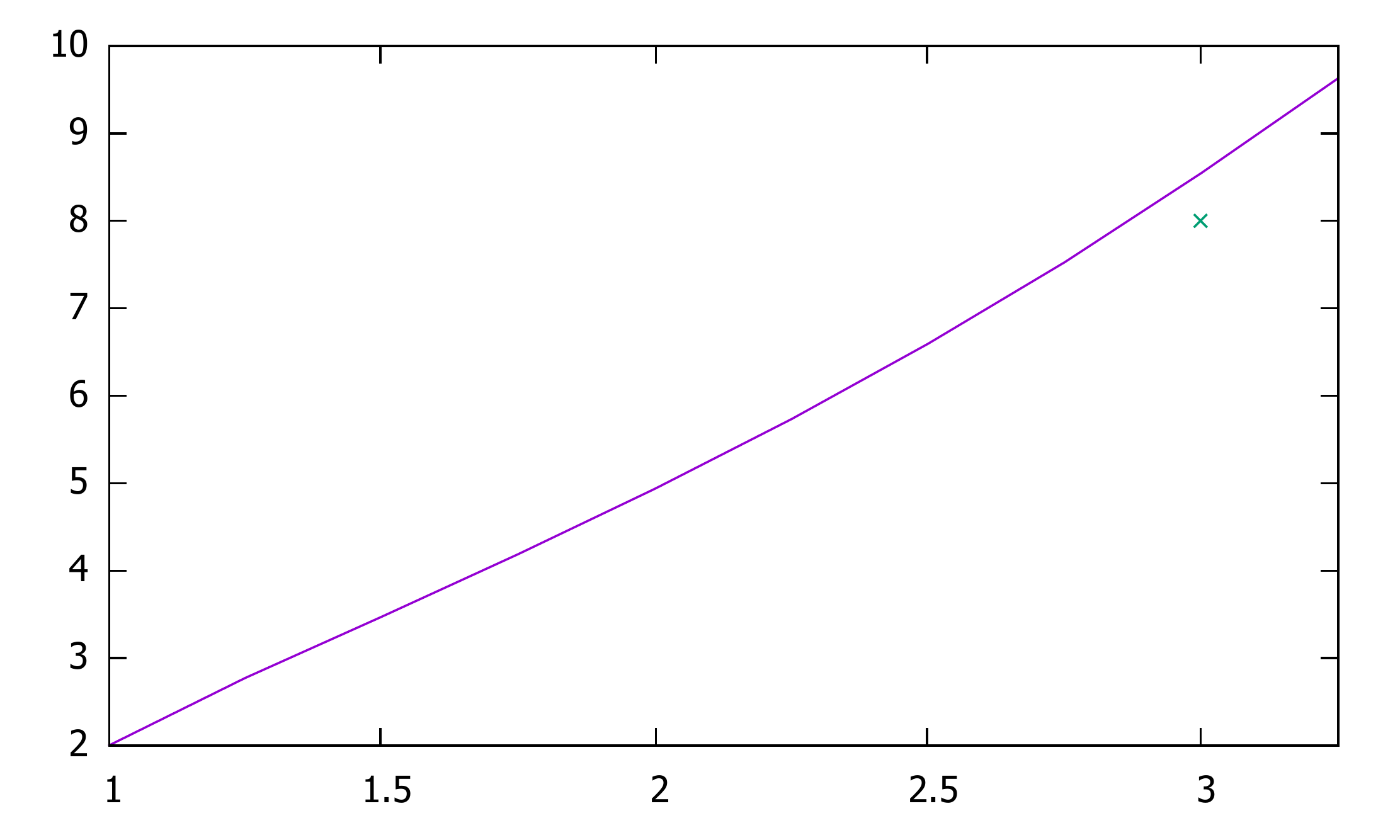}
	\end{center}
	\caption{Unitarity bound for $\Delta_{2}$ in four-dimensional $O(2)$ symmetric CFT as a function of $\Delta_{1}$. The point with green cross is the free massless Weyl fermion with the Pauli exclusion principle.}
	\label{fig:4do2}
\end{figure}

To assess the efficiency of the Pauli exclusion principle, let us compare it with the numerical conformal bootstrap bound with the $U(1)$ symmetry. In Fig \ref{fig:4do2} we have shown the bound. Our bound at $\Delta = 3.0$ is about $\Delta = 8.5$ (at $\Lambda = 41$). The bound is not yet converging, but it seems that the crude extrapolation of the $\Lambda \to \infty$ limit does not reach $\Delta = 8.0$. 

As in three-dimensions, it is an interesting question to ask if the slope becomes $2^{4/3}$, the number obtained from the asymptotic behavior of the conformal dimensions of charge $Q$ operators: $\Delta_Q = Q^{4/3}$ in four dimensions. The slope of the current bound is around 3 and larger than $2^{4/3}$, but this is necessarily so in order not to exclude the free fermion value of $\Delta_1 = 3.0$ and $\Delta_{2} = 8.0$. To be more conclusive, we need to study the larger $\Delta_1$, which becomes numerically harder within our approach.

\section{Discussions}
In this paper, we have studied the efficiency of the Pauli exclusion principle to obtain large anomalous dimensions. To assess the efficiency, we have compared  with the numerical conformal bootstrap bound. In two-dimensions, it can be most efficient and it saturates the bound. In higher dimensions, it is very close to the bound, but it may not saturate the numerical bound within the extrapolation we have attempted. 

Throughout the paper, we have mostly investigated free fermions, but the Pauli exclusion principle  can appear in the CFT spectrum with gauge symmetries. Certain gauge invariant composite operators cannot form a bound state without introducing further excitations. In supersymmetric gauge theories, it means that they give rise to non-trivial chiral ring relations.

In our discussions, we only studied free conformal field theories, so the Pauli exclusion principle can be implemented with no difficulty. However, in strongly coupled conformal field theories, the concept may become non-trivial. For example, in $\mathcal{N}=1$ supersymmetric  gauge theories in four-dimensions (say with $SU(N)$ gauge group), we may expect that the gaugino bilinear $S = \mathrm{Tr}W^\alpha  W_\alpha$ would show the Pauli exclusion principle so that $S^{N^2} =0$ because there are only $2(N^2-1)$ constituent fermions. However, this relation obtains non-perturbative quantum corrections and the simple Pauli exclusion principle does not apply in the strongly coupled gauge theories. In gapped theories, the chiral ring structure is $S^N=\Lambda^{3N}$, where $\Lambda$ is an appropriate confining scale but it is interesting to see what happens in the superconformal phase. 

Finally, we should admit that our brute-force way to find a functional to make a bound is probably not the most efficient way with larger $\Delta$. The use of the methods proposed in \cite{El-Showk:2016mxr} can be more suited for going with the flow to larger $\Delta$ and study the asymptotic behavior. A study of the large $\Delta$ sector is also important to understand the (AdS) black hole physics.\footnote{See e.g.\cite{Sen:2019lec}  for an analytic approach to the large $\Delta$ bound.}

\section*{Acknowledgements}
The author would like to thank  S.~Hellerman, D.~Orlando and S.~Reffert for discussions on the large charge CFTs and the conformal bootstrap bound. 
He also thanks S.~Rychkov for the correspondence. 
This work is in part supported by JSPS KAKENHI Grant Number 17K14301.

\end{document}